\documentclass[twocolumn,preprintnumbers,superscriptaddress,endnote,nofootinbib,prd]{revtex4}

\pdfoutput=1

\usepackage{graphicx}
\usepackage{dcolumn}   
\usepackage{amssymb}   
\usepackage{amsmath}

\newcommand{ \centeron }[2]{{\setbox0=\hbox{#1}\setbox1=\hbox{#2}\ifdim
                             \wd1>\wd0\kern.5\wd1\kern-.5\wd0\fi \copy0
                             \kern-.5\wd0\kern-.5\wd1\copy1\ifdim\wd0>\wd1
                             \kern.5\wd0\kern-.5\wd1\fi}}
\newcommand{ \ltap }{\>\centeron{\raise.35ex\hbox{$<$}}
                     {\lower.65ex\hbox{$\sim$}}\>}
\newcommand{ \gtap }{\>\centeron{\raise.35ex\hbox{$>$}}
                     {\lower.65ex\hbox{$\sim$}}\>}

\newcommand{ \slashchar }[1]{\setbox0=\hbox{$#1$}   
   \dimen0=\wd0                                     
   \setbox1=\hbox{/} \dimen1=\wd1                   
   \ifdim\dimen0>\dimen1                            
      \rlap{\hbox to \dimen0{\hfil/\hfil}}          
      #1                                            
   \else                                            
      \rlap{\hbox to \dimen1{\hfil$#1$\hfil}}       
      /                                             
   \fi}                                             %

\newcommand{\miss}{\slashchar{E}_T}
\usepackage[hypertex]{hyperref}

\newcommand{\lsim}{\lesssim}
\newcommand{\gsim}{\gtrsim}
\newcommand{\ra}{\rightarrow}

\begin{document}

\pagestyle{plain}

\title{\vspace*{-15mm}
\begin{flushright}
{\small {\rm FERMILAB-PUB-09-639-T}} \\
\end{flushright}
Discovering the Higgs Boson in New Physics Events using Jet Substructure}


\author{Graham D. Kribs}
\affiliation{Department of Physics, University of Oregon,
             Eugene, OR 97403}

\author{Adam Martin}
\affiliation{Theoretical Physics Department, Fermilab, Batavia, IL 60510}

\author{Tuhin S. Roy}
\affiliation{Department of Physics, University of Oregon,
             Eugene, OR 97403}

\author{Michael Spannowsky}
\affiliation{Department of Physics, University of Oregon,
             Eugene, OR 97403}


\begin{abstract}

We present a novel method to discover the Higgs boson in 
new physics event samples at the LHC\@.
Our technique applies to broad classes of models where
the Higgs has a significant branching fraction to $b\bar{b}$.
We exploit the recently developed techniques for discovering
a boosted Higgs using jet substructure.
Our requirements of new physics are quite general:
there must be features in the new physics event sample 
that allow a clean separation from standard model
background, and there should be Higgs bosons produced in 
association with the new physics.  We demonstrate
that this method superbly finds and identifies the lightest 
Higgs boson in the minimal supersymmetric standard model.
We focus on two case studies with a gravitino LSP, however, 
generalizations to other LSPs and to other models of new physics
are also briefly discussed. 
In some circumstances, discovery of the lightest Higgs is possible 
\emph{well before} conventional search strategies uncover 
convincing evidence.

\end{abstract}
\maketitle

\section{Introduction}

Discovering the Higgs boson is of the utmost importance 
for the LHC\@.  The most difficult discovery region is
when the Higgs mass is below about $130$ GeV,
where only rare decays, $h \ra \gamma\gamma$
and $h \ra \tau^+\tau^-$, have historically
been considered discovery modes.  

Recently, Ref.~\cite{Butterworth:2008iy} 
developed an intriguing   
technique to recover the dominant mode, $h \ra b\bar{b}$,
in the kinematic region where the Higgs is boosted.
A boosted particle that decays visibly in the detector 
leads to large transverse momentum (large $p_t$) where standard model 
backgrounds are smaller.  For boosted particles with hadronic decays, 
the large transverse momentum of the decay products can cause 
what would otherwise be two separate jets to appear in the detector
as a single ``fat jet''.  By ``fat jet'' we are referring to a 
jet with substructure consistent with that coming from a massive 
particle decay.  
For boosted particles that decay to $b\bar{b}$, the jet substructure
can contain two $b$-tagged subjets, which further reduces the 
background.  Amazingly, jet substructure techniques can achieve 
a theoretical significance up to $6$ for 
an integrated luminosity of $30$~fb$^{-1}$ at LHC  
\cite{Butterworth:2008iy}.
This has also very recently been validated by a realistic
simulation done by the ATLAS collaboration, achieving 
significance nearly this high \cite{atlas-study-vh}.

Jet substructure is itself insufficient to identify candidate 
Higgs events from the plethora of standard model backgrounds
in the absence of something triggerable.
Ref.~\cite{Butterworth:2008iy} relied on Higgs production 
in association with an electroweak gauge boson, $pp \ra Wh$, $pp \ra Zh$ 
with the $W$ or $Z$ decaying to leptons.  The lepton(s) provide 
crucial triggering for the candidate events and a powerful
discriminator against QCD backgrounds.

In this paper we consider the possibility that 
physics beyond the Standard Model, or simply ``new physics'',
is itself  
a substantial source of boosted Higgs decaying-to-$b\bar{b}$ production.
There are several reasons to think this is very promising: 
New physics can appear through QCD production, with a rate much
larger than electroweak production.
New physics can consist of particles significantly heavier than 
the Higgs boson, so that decays to the Higgs can, on average, 
lead to large boosts for every Higgs candidate event.  
Finally, new physics can have distinct triggerable collider signatures, 
some of which the LHC detectors are already designed 
to search, such as large missing energy $\miss$.

So, this paper assumes new physics has already been discovered and 
there is a distinctive enough signature of new physics 
to reliably separate out one or more ``new physics event samples''
that are reasonably pure. 
We then consider the region of high $p_t$, where we expect a 
near pristine new physics event sample with negligible standard model 
background.
This region has one very important advantage:  since detector 
resolution scales as $1/\sqrt{E}$, going to high $p_t$ necessarily
means we look into a region where detectors perform particularly well,
e.g., jet energy smearing is minimized. 
Within each event sample, we propose to carry out
a boosted-Higgs-via-jet-substructure search.  We find, in
some circumstances, the Higgs can be \emph{discovered}
within a jet-substructure-tagged search of new physics event samples  
well before conventional Higgs search analyses reach comparable 
statistical significance. 

We demonstrate the effectiveness of Higgs discovery using
supersymmetry as our example of new physics.  While our technique 
is completely general and model-independent, it is 
particularly well suited to the minimal supersymmetric standard 
model (MSSM).
This is because the lightest Higgs boson in the MSSM has
$m_h \lsim 125$~GeV (e.g.\ \cite{Carena:2000dp})
when superpartners are below about 1~TeV\@.  Secondly, the lightest
Higgs in supersymmetry decays dominantly to $b\bar{b}$, exactly
the mode we are searching for.
Finally, supersymmetry has a plethora of new particles that are 
at or beyond the electroweak scale.  Some of these particles 
can decay into the Higgs, and in some instances, with a fairly 
large branching fraction. 

Nevertheless, we stress that our methodology is model-independent.
For every sample of clean (nearly background-free) new physics events, 
a boosted Higgs via jet substructure should be performed.
Whether the search is successful depends on a convolution
of the new physics scenario as well as the detector performance.
Important details of the new physics model include the 
total cross section of new physics,
the fraction of new physics produced that can be cleanly separated
from standard model backgrounds, 
the fraction of this sample that has Higgs bosons resulting from 
new heavy particle decays, and the fraction of these Higgs bosons 
that are boosted.
Important detector performance details include the $b$-tag efficiency,
which includes tagging a jet as well as subjets, the jet energy 
resolution, fake rates, and so on.

\section{Boosted Higgs}

A boosted Higgs boson has high transverse momenta $p_t \gg m_h$.
When the Higgs decays to $b\bar{b}$, this high transverse momenta 
causes the $b$-jets to be highly collimated.  
Conventional search strategies to identify the Higgs through
the reconstruction of two separate singly $b$-tagged jets 
generally fails since it is much more likely for the 
$b$-jets to be merged into a single jet.  Going to smaller
cone size would seem prudent, except that this has been shown
to give poor mass resolution \cite{Seymour:1993mx}. 

Instead, we exploit the recently developed technique to identify 
subjets within a ``fat jet'' consistent with the decay of a 
Higgs to $b\bar{b}$ \cite{Butterworth:2008iy}.
Identifying subjets inside a fat jet that resulted from the decay 
of a massive particle is not straightforward.  
Jet mass, determined by some algorithmic prescription applied
to the subjets, is one indicator.
However, the distribution that results from ordinary QCD production 
still has a long tail into high jet masses.
For a jet with transverse momentum $p_t$, jet mass $m_j$, 
and cone size $R^2 = \Delta \eta^2 + \Delta \phi^2$,
the leading order differential QCD jet mass distribution 
goes as \cite{Almeida:2008yp,Almeida:2008tp} 
\begin{equation}
  \label{eq:jet-mass-dist}
  \frac{ d \sigma \left( R \right) }{d p_t d m_j} \sim
  \frac{\alpha_s C_i}{\pi m_j^2} \left( \ln \frac{R^2 p_t^2}{m_j^2}
  + \mathcal{O}\left( 1 \right)  \right) \; .
\end{equation}
The challenge is thus to reduce the QCD jet background without losing 
significantly in mass resolution. 
Further, when a jet with substructure is identified, we also need 
to determine the ``heavy particle neighborhood'' -- the region 
to which 
QCD radiation from the Higgs decay products 
is expected to be confined.

Analysis of jet substructure has received considerable attention. 
Distinct algorithms have been proposed to identify
Higgs decaying to $b\bar{b}$ \cite{Butterworth:2008iy,Plehn:2009rk},
fully hadronic decays of
top \cite{Thaler:2008ju,Kaplan:2008ie,Brooijmans:2008zz,Plehn:2009rk},
and even neutralinos decaying to 
three quarks \cite{Butterworth:2007ke,Butterworth:2009qa}.  
Refs.~\cite{Ellis:2009su,Ellis:2009me,Krohn:2009th}
have also recently introduced 
a more general ``pruning'' procedure based on jet substructure to
more easily discover heavy particles.
Our work employs a modified version of the iterative decomposition
algorithm introduced by Ref.~\cite{Butterworth:2008iy}, which uses 
an inclusive, longitudinally invariant Cambridge/Aachen (C/A)
algorithm \cite{Dokshitzer:1997in,Wobisch:1998wt,Wobisch:2000dk}.

\section{Jet Substructure Algorithm}

The starting point to test our algorithm, both for new physics 
and SM background processes, is a set of final (post-showering 
and hadronization) particles.  
We generate signal events using Pythia v6.4~\cite{Sjostrand:2006za},
while the background events are first generated at parton-level
using ALPGENv13~\cite{Mangano:2002ea}.  We use PYTHIA v6.4 for showering
and hadronization of all events. We also use the 
ATLAS tune~\cite{Buttar:2004iy} in Pythia to model the underlying event.  
We do not perform any detector simulation or smearing of jets. 
A realistic ATLAS/CMS specific search in the spirit of 
Ref.~\cite{atlas-study-vh} is beyond the scope of this work. 
However, since high $p_t$ jets result in a large amount of energy 
deposited in the calorimeter cells where energy resolution is
excellent, we do not expect smearing to significantly modify our
results.

We group the hadronic output of Pythia into ``cells'' of size  
$\Delta \eta \times \Delta \phi = 0.1 \times 0.1$\@. 
We sum the four momentum of all particles in each
cell and rescale the resulting three-momentum such as to make the cells
massless \cite{Thaler:2008ju}.
If the cell energy is bigger than $1$~GeV, the cells
become the inputs to the jet algorithm.  We use the inclusive C/A
algorithm as implemented in FastJet~\cite{Cacciari:2005hq} to
cluster the input cells in jets with $R = 1.2$\@.  
As we are trying to identify the Higgs through its decay to bottom quarks, 
the $b$-tag efficiency is paramount.
For simplicity we work with a flat $60\%$ acceptance, 
with a corresponding fake rate of $2\%$\@.
Our algorithm is as follows:

\begin{enumerate}

\item The decomposition procedure starts with a $b$-tagged jet $j$. 
  After undoing its last stage of clustering, the two subjets 
  $j_1$ and $j_2$ are labeled such that $m_{j_1} > m_{j_2}$\@. 

\item Following Ref.~\cite{Butterworth:2008iy}, 
  subjets are checked for the existence of a significant mass drop 
  ($m_{j_1} < \mu\, m_j$) as well as non-existence of an asymmetry defined by
  $y = \frac{\text{min}\left( p_{t_{j1}}^2,  p_{t_{j2}}^2 \right) }{m_j^2}
  \Delta R^2_{j_1,j_2} > y_\text{cut}$.  We use 
  $\mu = 0.68$ and $y_\text{cut} = (0.3)^2$ identical to 
  Ref.~\cite{Butterworth:2008iy}.  
  Both subjets are required to be $b$-tagged and the $p_t$ of the
  daughter jet $j$ greater than $30$~GeV\@. 
  If these conditions are satisfied, this stage of clustering (say, $i$-th) 
  is recorded and then the following is calculated:
  \begin{equation}
    \label{eq:discriminant}
    S_i = \frac{\text{min}\left( p_{t_{j_1}}^2,  p_{t_{j_2}}^2 \right) }
       {\left( p_{t_{j_1}} +  p_{t_{j_2}} \right)^2 } \Delta R_{j_1
         j_2} \; .
  \end{equation}
  The quantity $S_i$ is an indicator of the similarity of the 
  two subjets and is weighted by their separation $\Delta R_{j_1 j_2}$. 

\item Replace $j$ by $j_1$ and repeat from step~$1$ as long as $j$
  has further subjets.

\item Select the stage of clustering for which $S_i$ is the
  largest.  We anticipate that the two $b$-tagged subjets, at this stage,
  are most likely to have originated from Higgs decay since they are 
  more likely to be similar to each other.  If the two C/A $b$-tagged
  subjets 
  originate from Higgs decay, the subjets with opening 
  angle $\Delta R_{j_1 j_2}$ should contain all the perturbative
  radiation from the $b\bar{b}$ system by virtue of angular
  ordering \cite{Bassetto:1984ik}.  However, the subjets still tend 
  to include too much contamination from underlying events.  We then
  filter \cite{Butterworth:2008iy} the events:
  we cluster the jet constituents again using a finer angular scale 
  specific to the jet [we use, $\text{min} \left( R_{bb}/2, 0.3\right)$]
  and retain only the three hardest components $(b \bar{b} g)$.
  Finally, we combine the three subjets and call the result
  a ``candidate Higgs jet''.

\end{enumerate}

\section{Supersymmetry as an Example}

We now demonstrate the effectiveness of our algorithm
for one particularly interesting example of new physics, 
the minimal supersymmetric standard model.
One of the most promising regions of parameter space with
large associated Higgs production occurs when 
the next-to-lightest supersymmetric particle (NLSP) 
is a neutral Higgsino and the lightest supersymmetric particle
(LSP) is the gravitino \cite{Matchev:1999ft}.

In this paper we consider two sets of supersymmetric parameters,
given in App.~\ref{spectrum-app}.
The first set, ``Point 1'', follows Ref.~\cite{Matchev:1999ft} with a 
Higgsino NLSP and gravitino LSP\@.  Point 1 is, not surprisingly, 
quite favorable to our analysis, with nearly one Higgs boson 
produced in association with every pair of superpartners produced.  
It represents the incredible power of this analysis to reach 
high significance of light Higgs discovery with relatively low 
luminosity at LHC.  

The second set, ``Point 2'', still contains a gravitino LSP,
but a bino LSP\@.  This point is seemingly much less favorable
to our analysis, with only about 4\% of superpartner pair production 
containing a Higgs boson from NLSP decay.  Nevertheless, we will
see that the Higgs signal can even be found here. 
With visible signals with just $4\%$ of new physics production, 
our methodology clearly has the potential to much wider applicability 
within the vast supersymmetric parameter space 
(already in progress in Ref.~\cite{preparation})
as well as applied to other models of new physics.  

At first glance one might think that these two Points seems rather
specific and not particularly general, even for supersymmetry.
However, a gravitino LSP is an automatic consequence of a 
low supersymmetry breaking scale.  
LEP II bounds from the search for Higgsino-like charginos 
implies $\mu \gsim 100$ GeV, and Tevatron bounds from the 
lack of observation of inclusive $\gamma\gamma + \miss$ 
implies $M_1,\mu$ somewhat heavier than 
this \cite{Ambrosanio:1996jn,Baer:2000pe,Meade:2009qv}.
The decay of a NLSP into the Higgs is thus essentially
always kinematically open.
More typical neutralino mixing suggests that the NLSP is partly
Higgsino and partly bino, the mixture determined by the ratio
of the Higgsino mass to bino mass, $\mu/M_1$.
This implies the NLSP could decay into $h$, $Z$, or $\gamma$ 
depending on supersymmetric parameters.  Our choices of
``Point 1'' and ``Point 2'', provide two 
particularly interesting examples into this parameter space.

In both Points 1 and 2, several clean new physics event samples exist. 
The most distinctive event sample at the LHC is 
$\gamma\gamma + \miss$, which would provide the 
clearest first evidence for physics beyond the standard model.
Only relatively mild cuts on the photon energy and the total
missing transverse energy are necessary to eliminate the SM background.
Right behind this discovery will be a larger sample of 
$\gamma + \miss$ events, and with suitable cuts,
that we explain in detail below, provide another clean new physics
event sample with associated Higgs production.

Colored superpartners are taken to be 750 GeV for both Points.
This is well outside the sensitivity of Tevatron searches and yet
well within LHC capabilities with order fb$^{-1}$ integrated
luminosity.  Our goal is find Higgs bosons
produced from NLSP decay, and so we are interested in 
inclusive superpartner production,
\begin{equation}
  \label{eq:process}
  p + p \ \ra \ \chi_1 + \chi_1 + \text{anything}
   \ \ra \ h + \gamma  + X + \miss \; , 
\end{equation}
where $X$ can be jets and/or leptons resulting from the
superpartner decay chains.  We choose to focus on $\gamma + \miss$
as the primary discrimant of signal from background, but other
signal samples such as $Z + \miss$ may well be more appropriate
in other regions of parameter space.
For both of the Points (c.f.\ App.~\ref{spectrum-app}), 
the dominant source of superpartner production is first and 
second generation squark production.  To gain some idea of the amount of
signal we are working with for Points (1,2): 
the branching fraction to $h + \gamma$ is about ($16\%$,$4\%$);
the fraction of boosted Higgs bosons is about ($40\%$,$50\%$);
this leads to a cross section for boosted $h + \gamma$
of about ($0.13$,$0.07$) pb. 

To simulate the supersymmetric signal, we use 
Pythia v6.4 \cite{Sjostrand:2006za} to generate parton 
level events with subsequent showering and hadronization. 
We assume the LHC is operating at $\sqrt{s} = 14$~TeV,
however our results follow quite well for lower energies
that we have also simulated, including $\sqrt{s} = 10$~TeV\@.  
We take leading order cross sections for squark and gluino production,
however note that the k-factors can be sizeable \cite{Beenakker:1996ch}.  
We select events that have at least one hard and isolated photon,
at least one $b$-jet with high $p_t$, and have large missing energy:
\begin{equation}
  \label{eq:cut-level1}
  \miss > 100~\text{GeV} \qquad  p_t^\gamma > 100~\text{GeV}
  \qquad  p_t^b  > 200~\text{GeV} \; .
\end{equation}
For simplicity, we have taken the cuts to be identical for
both Points despite the fact that further optimization in
significance can be achieved by tuning the cuts.
More details of the effects of varying cuts will be 
presented in Ref.~\cite{preparation}.

\begin{figure*}[t]
\begin{center}
\includegraphics[width=0.48\textwidth]{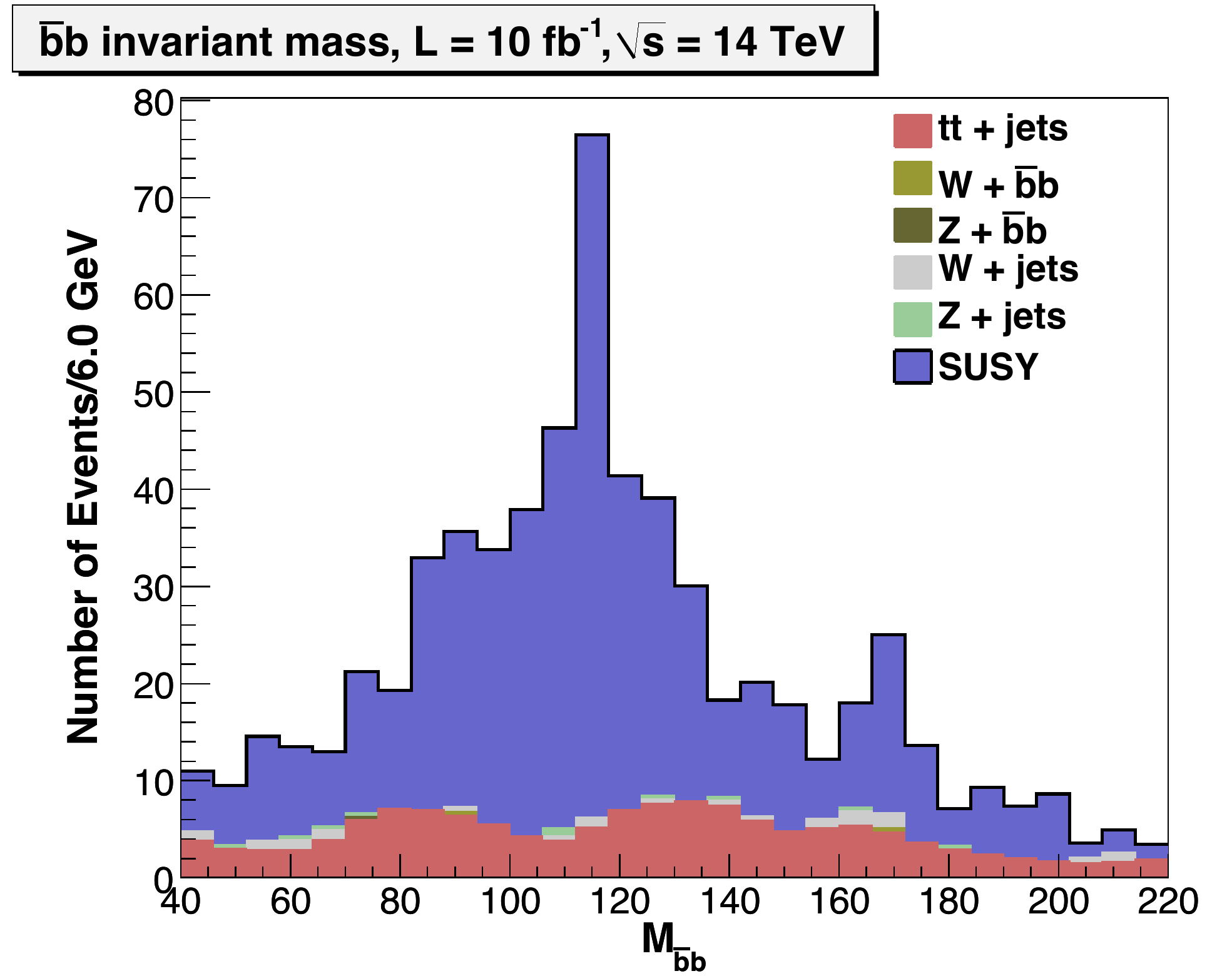}
\hspace*{0.02\textwidth}
\includegraphics[width=0.48\textwidth]{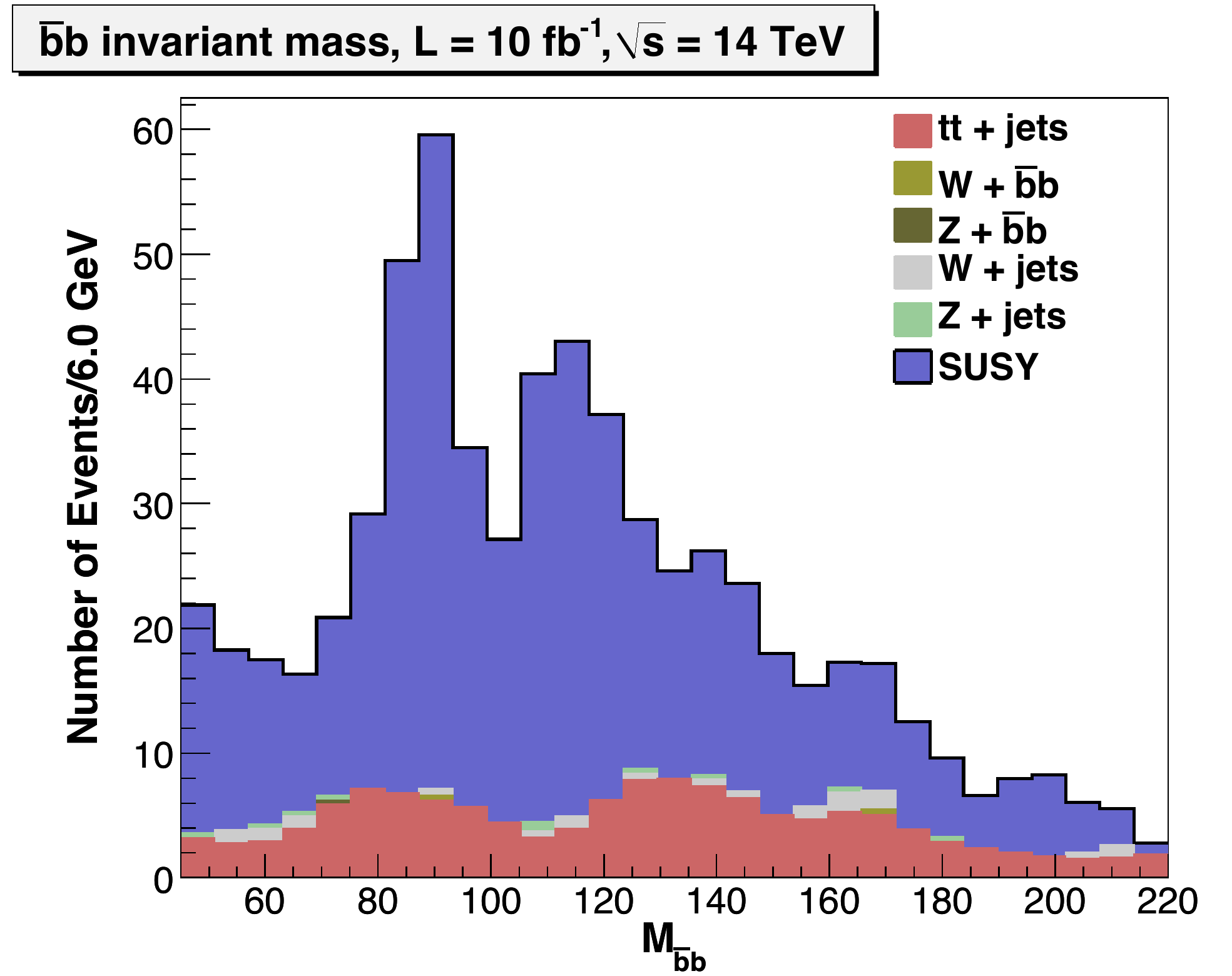}
\vspace*{-3mm}
\end{center}
\caption{The Higgs peak in the $b\bar{b}$ invariant mass distribution 
is easily resolved above the supersymmetric and standard model backgrounds,
and well separated from the $Z$ peak, using our jet substructure algorithm.  
The left figure (a) corresponds
to ``Point 1'' (Higgsino NLSP) while the right figure (b) corresponds to 
``Point 2'' (Bino NLSP).  See text for details.  (color online)}
\label{fig:results}
\end{figure*}

To be conservative, we take the photon identification 
efficiency to be $80\%$, with a jet-fakes-photon rate of 
$0.1\%$ \footnote{A jet rejection factor of $\simeq 0.05\%$
  can actually be obtained in ATLAS with photon
  identification efficiency of $80\%$~\cite{atlas-study-gamma}.}. 
We then run our Higgs-finding algorithm on the event sample that
survives the cuts listed in Eq.~(\ref{eq:cut-level1}). 
Roughly one in four events with a Higgs boson pass this cut.
The output of the algorithm contains candidate Higgs jets.
We show the result of our algorithm
in Fig.~\ref{fig:results}.

In Fig.~\ref{fig:results}, 
the shaded region labeled SUSY contains many ``candidate Higgs jets'' 
that are not, in fact, from Higgs decay.  
Instead, we must look for features in the $b\bar{b}$ invariant 
mass distribution, and it is here that one of the great advantages
of using jet substructure becomes manifest:  The clear identification
of a Higgs peak in both Points!  In both Points, neutralinos that decay 
to $Z(\ra b\bar{b}) + \tilde G$ also pass our cuts.  
The $Z$ peak is easily visible in Point 2, while in Point 1
it forms merely an extended shoulder towards lower invariant
jet mass.  For Point 2, both the $Z$ and $h$ peaks are clearly evident, 
illustrating the importance of the high energy resolution attained 
using jet substructure.\footnote{A conventional analysis relying
on reconstructing two separate $b$-jets completely misses 
the Higgs peak in Point 2.}
The smaller feature near $m_{b\bar{b}} \sim 160$-$180$~GeV 
in both plots is due to hadronic top decay where either a mistag 
or Cabbibo-suppressed $W^{\pm}$ decay resulted in two $b$-tagged 
jets in the same initial fat jet.

In this analysis the major Standard Model background is 
$t\bar{t} + \mathrm{jets}$ 
where one of the jets fakes an isolated photon.
The $t\bar{t} + \mathrm{jets}$ background is modeled 
as the sum of leading order $t\bar{t} + 0$, $+1$, $+2$ jet
samples with total cross section 
$1110$~pb.\footnote{The ALPGEN samples were generated 
without jet-parton matching.  However we rescale the total 
cross section to a sample of CKKW matched MadGraph \cite{Alwall:2007st}  
$t\bar{t} +0,1,2$~jets.}
Further reducible and irreducible 
Standard Model backgrounds, 
such as $W +\gamma+\mathrm{jet}$ \cite{Campanario:2009um},  
are negligible after applying Eq.~(\ref{eq:cut-level1}) and a $b$-tag. 
The $W/Z + \text{jets}$ refers to $+3$ or more jets; 
we find that exclusive processes with two or fewer jets 
do not contain enough hard objects to fake our signal.

We can make a crude estimate of the significance of the Higgs peak 
for each Point where all SM events and continuum supersymmetric events 
are counted as background.  The supersymmetric continuum is simply 
derived by connecting the histograms 
between $-2(-1)$~to~$+3(+2)$ bins on either side of 
the Higgs peak in Point 1 (2) with a straight line; 
all events in the lower trapezoid 
are the background, while the event above are the Higgs signal. 
This procedure yields a significance 
$\mathcal S = S/\sqrt B \simeq (9,5)$ 
for Points (1,2) as shown in Fig.~\ref{fig:results}.
Adjusting the cuts, either by hardening the photon $p_T$ 
requirement or by adding a cut on the scalar $p_T$ sum, 
$H_T \gtrsim 600$~GeV, can increase this slightly.  
Remarkably, with as little as a few fb$^{-1}$, a light
Higgs with mass $115$~GeV could be discovered in the
new physics event sample if the branching fraction to Higgs
bosons is order one.   
Similarly, with just 
$10$~fb$^{-1}$, strong evidence of a light Higgs can 
appear in new physics scenarios where only about $4\%$ of
the new physics event sample contains a Higgs at all!

By comparison, conventional cut-based searches for Higgs bosons in 
$\gamma\gamma, \gamma\gamma + \text{jets}$ require integrated luminosity 
$\sim 30$~fb$^{-1}$~\cite{Aad:2009wy,deRoeck:942733}. 
By combining channels and using sophisticated optimization, 
the discovery luminosity can be brought down closer to 
$10$~fb$^{-1}$~\cite{deRoeck:942733}, still well above the result 
of our simple analysis.  Obviously this comparison is blatantly
unfair because the ATLAS/CMS studies include a complete
detector simulation.  Nevertheless, we do think it is intriguing
that comparably high levels of (theoretically-derived) significance 
are possible with this methodology.

\section{Discussion}

We have demonstrated that Higgs bosons produced copiously
with distinctive superpartner production can be easily
found with a jet substructure analysis.  Moreover, we have
also shown that a Higgs feature in the jet invariant mass distribution
can be seen with only a few percent of superpartner production 
containing a Higgs.  
This goes far beyond what conventional supersymmetric analyses, 
e.g.\ Refs.~\cite{Datta:2003iz,Carena:2005ek,deCampos:2008ic}
have achieved. 
We are now well underway simulating even more general MSSM models 
with or \emph{without} a gravitino LSP, and our preliminary results
suggest we can still easily identify the Higgs in the new physics 
event sample.  Details will be presented in Ref.~\cite{preparation}.

We cannot overemphasize enough that our study using supersymmetry
as new physics was nothing more than a convenient example, 
and our methodology can be applied to many models of new physics 
observable at the LHC\@.
Indeed, any new physics model that has a distinctive signal 
(identified clearly from SM background) as well as associated 
Higgs production 
(with the Higgs decaying some of the time to $b\bar{b}$)
can benefit from our proposal.  It would be very interesting
to consider new physics event samples from little Higgs models or
universal extra dimension 
models with associated Higgs production.

Finally, given the remarkable progress of dark matter direct detection
experiments, there is a huge opportunity for nuclear recoils
to be seen in the near future.  Many models of new physics with
dark matter have a leading-order elastic-scattering cross section
dominated by the exchange of a light Higgs boson.  
If dark matter direct detection is observed 
consistent with elastic scattering, the techniques 
and methodology presented here becomes of exceptional
importance to rapidly determine the connections between new physics,
Higgs physics, and dark matter.  Perhaps the most exciting prospect 
is that this technique could unite the origin of electroweak symmetry 
breaking with new physics with dark matter!

\acknowledgments

MS thanks T.~Plehn and G.~Salam for helpful discussions.
GDK and TSR thank the Galileo Galilei Institute for Theoretical 
Physics for hospitality where part of this work was completed.
This work was supported in part by the US Department of Energy 
under contract number DE-FG02-96ER40969 (GDK, TSR, MS).
AM is supported by Fermilab operated by Fermi Research Alliance, 
LLC under contract number DE-AC02-07CH11359 with the 
US Department of Energy.


\appendix

\section{Spectrum}
\label{spectrum-app}

The parameters, part of the spectrum, and some relevant
collider information used in this analysis 
is presented in Table~\ref{table:spectra}.  
The spectrum was computed
with SUSPECT2~\cite{Djouadi:2002ze}. 
The spectra share a moderate $\tan\beta$, large $m_A$ and large $a_t$ -- 
resulting in heavy, nearly degenerate $H^0, A^0$ and $H^{\pm}$ 
at $\sim 1$~TeV and a large $\tilde{t}_1 - \tilde{t}_2$ splitting.
In Point 1 the lightest neutralino is primarily Higgsino, 
while in Point 2 it is predominantly bino. 
In addition to the mass eigenstates shown above, both spectra contain 
heavy charginos and neutralinos at $\sim M_2$. 
The total inclusive supersymmetric particle production cross section
at LHC to leading order is given by $\sigma_{\text{tot}}$.
The sleptons, $m_{\tilde{\ell}} = 500$~GeV in Point 1 and 
$m_{\tilde{\ell}} = 1$~TeV in Point 2, play essentially no role. 
The final parameter the two spectra share is the gravitino mass; 
we use a light gravitino ($1$~eV) for a prompt decay.

\begin{table}[h]
\squeezetable
 \begin{tabular}[c]{c|cc}
 & Point~1 & Point~2 \\ \hline
 $\tan\beta$                       & $10$       & $10$       \\
 $M_1$                             & $300$~GeV  & $240$~GeV  \\
 $M_2$                             & $600$~GeV  &   $1$~TeV  \\
 $M_3$                             &   $2$~TeV  &   $1$~TeV  \\
 $\mu$                             & $-250$~GeV & $260$~GeV  \\
 $m_A$                             & $1$~TeV    &   $1$~TeV  \\
 $a_t$                             & $-920$~GeV & $-920$~GeV \\
 $m_{\tilde{q}}$                   & $750$~GeV  & $750$~GeV  \\ \hline
 $m_h$                             & $115$~GeV  & $116$~GeV  \\
 $\chi_1$                          & $235$~GeV  & $213$~GeV  \\
 $\chi_2$                          & $260$~GeV  & $268$~GeV  \\
 $\chi_1^{+}$                      & $250$~GeV  & $261$~GeV  \\ \hline
 $\sigma_{\text{tot}}$ ($\sqrt{s} = 14$~TeV) & $2.3$~pb & $3.7$~pb \\ 
 $BR(\tilde{\chi}_1^0 \ra h\tilde{G})$       & $0.30$   & $0.02$   \\ 
 $BR(\tilde{\chi}_1^0 \ra Z\tilde{G})$       & $0.35$   & $0.21$   \\  
 $BR(\tilde{\chi}_1^0 \ra \gamma\tilde{G})$  & $0.35$   & $0.77$       
 \end{tabular}
\label{table:spectra}
\caption{The parameters, part of the spectrum, and some relevant
collider information for the two Points used in this analysis.
See text for details.}
\end{table}


\end{document}